%% file: main.tex
\newif\ifanonymous
\anonymousfalse

\newif\ifshortpaper
\shortpapertrue

\ifanonymous
\documentclass[review]{vgtc}
\else
\documentclass{vgtc}
\fi

\usepackage{mathptmx}                        
\usepackage{times}                           
\usepackage[T1]{fontenc}
\usepackage{graphicx}
\usepackage{amsmath}
\usepackage{microtype}
\usepackage{xspace}
\usepackage{enumitem}
\usepackage{multirow}
\usepackage{booktabs}
\usepackage{float}
\usepackage{balance}
\usepackage{comment}
\usepackage[colorinlistoftodos,prependcaption,textsize=small]{todonotes}

\ifshortpaper
\setlist[itemize]{leftmargin=*,topsep=3pt,itemsep=2pt,parsep=0pt,partopsep=0pt}
\setlist[enumerate]{leftmargin=*,topsep=3pt,itemsep=2pt,parsep=0pt,partopsep=0pt}
\fi

\usepackage[pdftex]{hyperref}
\hypersetup{
  pdftitle  = {ImputeViz: A Visual Analytics Dashboard for Diagnosing Missing Data and Comparing Imputation Methods},
  pdfauthor = {\ifanonymous Anonymous IEEE VIS 2026 submission\else ImputeViz Authors\fi},
  bookmarksnumbered,
  pdfstartview={FitH},
  breaklinks=true,
  colorlinks=false,
  hidelinks
}

\onlineid{1359}                                 
\vgtccategory{Research}
\vgtcpapertype{application/design study}

\newcommand{\gknn}{gKNN\xspace}
\newcommand{\mice}{MICE\xspace}
\newcommand{\tool}{ImputeViz\xspace}

\title{ImputeViz: A Visual Analytics Dashboard for Diagnosing Missing Data and Comparing Imputation Methods}

  \author{Aitik Dandapat\thanks{e-mail: adandapat@cs.stonybrook.edu}\\
          \scriptsize Stony Brook University, New York.
  \and   Lalith Punepalle Raveendrareddy\thanks{e-mail: lpunepallera@cs.stonybrook.edu}\\
          \scriptsize Stony Brook University, New York.
  \and   Mithilesh Singh\thanks{e-mail: mkssingh@cs.stonybrook.edu}\\
          \scriptsize Stony Brook University, New York.
  \and   Klaus Mueller\thanks{e-mail: mueller@cs.stonybrook.edu}\\
          \scriptsize Stony Brook University, New York.}

  \authorfooter{
    \item Aitik Dandapat is with Stony Brook University. E-mail: adandapat@cs.stonybrook.edu.
    \item Lalith Punepalle Raveendrareddy is with Stony Brook University. E-mail: lpunepallera@cs.stonybrook.edu.
    \item Mithilesh Singh is with Stony Brook University. E-mail: mkssingh@cs.stonybrook.edu.
    \item Klaus Mueller is with Stony Brook University. E-mail: mueller@cs.stonybrook.edu.
  }

\ifanonymous
\author{Anonymous Author(s)}
\authorfooter{
  \item Author information omitted for double-blind review.
}
\else
\author{
  Aitik Dandapat\thanks{e-mail: adandapat@cs.stonybrook.edu}\\
  \scriptsize Stony Brook University
\and
  Lalith Punepalle Raveendrareddy\thanks{e-mail: lpunepallera@cs.stonybrook.edu}\\
  \scriptsize Stony Brook University
\and
  Mithilesh Kumar Singh\thanks{e-mail: mkssingh@cs.stonybrook.edu}\\
  \scriptsize Stony Brook University
\and
  Klaus Mueller\thanks{e-mail: mueller@cs.stonybrook.edu}\\
  \scriptsize Stony Brook University
}
\authorfooter{
  \item Aitik Dandapat is with Stony Brook University. E-mail: adandapat@cs.stonybrook.edu.
  \item Klaus Mueller is with Stony Brook University. E-mail: mueller@cs.stonybrook.edu.
}
\fi

\abstract{%
Missing data is a persistent obstacle in scientific, social science, and public health research, often biasing analyses and placing accountability on analysts for how they handle missing values. We introduce \textsc{ImputeViz}, an integrated visual analytics dashboard that supports diagnosing missingness, configuring imputation models, and evaluating results. The system brings together widely used methods---including MICE, Random Forest, XGBoost, and kNN---within an interactive environment that makes missingness patterns explicit. To support geospatial reasoning, we introduce gKNN, a geographically informed kNN variant that blends socio-economic and spatial distances and exposes \emph{donor contributions}, enabling provenance-based visual accountability by showing which regions drive each estimate. Our primary contribution is a method-agnostic visual analytics environment that makes cross-method comparison a first-class visual task and integrates gKNN alongside standard methods. Coordinated views reveal missingness structure through heatmaps, co-missingness summaries, and distributional diagnostics that help analysts reason about missingness patterns (MCAR/MAR) and cases where missingness may be non-random (MNAR). Users can compare and tune models and interrogate results via distributional overlays, a \emph{Method Comparison Summary} reporting MAE, RMSE, $\Delta$RMSE, and runtime for each algorithm on the current target and mask, along with variable-level discrepancy views. Cached per-method results and locked axis scales reduce cognitive overhead from shifting ranges during method switching. These comparisons highlight where methods disagree, which variables are sensitive, and how imputation choices affect downstream summaries. Case studies demonstrate how \textsc{ImputeViz} helps analysts select effective strategies, surface sensitive variables, and assess model robustness.%
}


\begin{document}

\firstsection{Introduction}

\maketitle

\ifshortpaper
\input{ieee_vis_short_content}
\fi
\ifanonymous\else
\section*{Acknowledgments}
Portions of this manuscript were refined with assistance from ChatGPT-5.1 to improve clarity, structure, and wording. All conceptual contributions, system design decisions, analyses, and substantive content originated from the authors.
\fi

\balance
\bibliographystyle{abbrv-doi}
\bibliography{gknn}

\end{document}

%% file: ieee_vis_short_content.tex
\label{sec:introduction}
Missing data remains a persistent obstacle in scientific and public-health analysis. In county-level public-health surveillance, privacy suppression and sparse reporting are often systematic: missing entries cluster in low-count regions, which can bias downstream trend analysis and resource allocation if imputed poorly. At the same time, many applied workflows treat imputation as a black-box preprocessing step and provide limited support for cross-method visual comparison or for inspecting why a value was imputed in a particular way -- often forcing analysts to compare models sequentially and mentally reconcile shifting scales and parameters across runs.

We present \tool, a visual analytics dashboard that supports the full imputation workflow: diagnosing missingness, selecting and tuning models, comparing outcomes, and inspecting provenance. The system integrates common imputers (MICE, Random Forest, XGBoost, linear/kNN baselines) and a geospatially informed method (\gknn) that blends socioeconomic and spatial distance. Rather than promoting one model universally, \tool is designed to make method choice auditable through linked diagnostics, stable-scale comparisons, and quantitative summaries.

Our contributions are threefold: (1) a visual analytics workflow for missingness diagnosis and cross-method comparison under stable-scale linked views; (2) a benchmarking backend that caches per-method runs and enforces consistent holdout-based evaluation for fair comparison; and (3) \gknn, a geographically informed kNN variant with donor provenance that can be interrogated directly in the interface.

\section{Related Work}
Imputation approaches include chained equations (MICE), tree ensembles, and instance-based methods~\cite{VanBuuren2018MICE, Breiman2001RandomForest, Chen2016XGBoost, Cover1967KNN}. Spatial extensions of nearest-neighbor imputation can better exploit locality when outcomes are geographically autocorrelated~\cite{baker2014missing}. Our \gknn builds on this line with a weighted blend of socioeconomic and geographic distance learned via Bayesian optimization~\cite{snoek2012practical}.

Prior VA systems support missingness diagnosis and imputation validation in cohort and healthcare settings, including work by Alemzadeh et al.~\cite{alemzadeh2017visual, alemzadeh2020visual}. Related efforts also study visualization support for imputation workflows and model-based imputation in interactive analysis~\cite{templ2022visualization, yeon2022visual}. \tool complements these by enabling multi-method execution and cross-method comparison in one interface via cached runs, consistent holdout evaluation, locked visual scales, and donor-level provenance for \gknn.

Visual analytics research emphasizes integrating computational models with interactive reasoning~\cite{Endert2017}. Prior systems in public-health VA emphasize interactive exploration and the role of uncertainty and trust in analysis~\cite{sankaran2016opioid, Carroll2014HealthVis, Sacha2016UncertaintyTrust}; in our setting, uncertainty estimation is currently computed offline due to cost and reported as supporting evidence in our evaluation.


\begin{figure*}[t]
  \centering
  \includegraphics[width=0.9\linewidth]{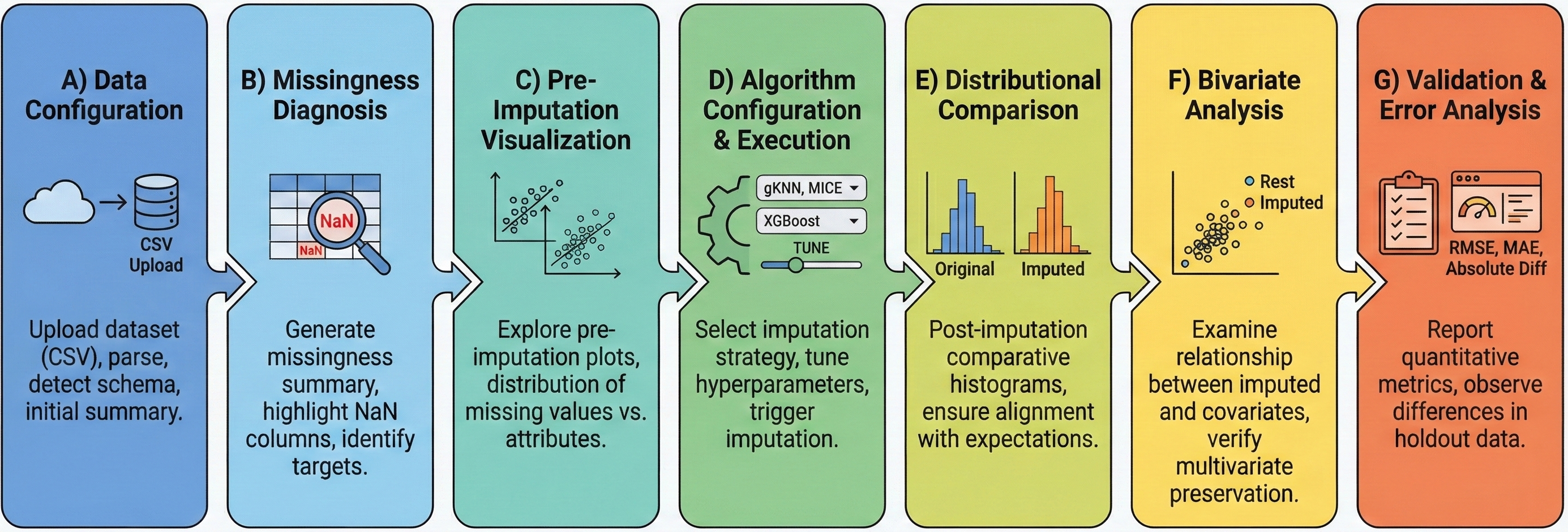}
  \caption{Seven-stage \tool workflow from data upload and missingness diagnosis to model execution, visual comparison, and validation.}
  \label{fig:workflow_short}
\end{figure*}

\section{Data}
We evaluate on two datasets. First, county-level U.S.\ prescription-opioid mortality (2000--2022) from CDC WONDER~\cite{cdcWonderDrugPoisoning}, joined with ACS socioeconomic covariates~\cite{acsPopulation} and county geometries~\cite{tigerShapefiles}. Missingness is structured rather than MCAR~\cite{LittleRubin2019} because suppression concentrates in low-count counties and often forms spatial clusters. Second, a non-geographic telecom churn dataset~\cite{telcoHF} (212 rows after filtering), where we impute the continuous \texttt{Offer} target from numeric covariates.

\section{Design Goals}
Based on analyst feedback, we target four design goals that guide both algorithm integration and interface decisions.

\paragraph*{DG1: Make missingness legible.}
Analysts need immediate visibility into where missing values occur, whether they cluster by feature or region, and whether missingness correlates with plausible covariates. The system therefore prioritizes missingness summaries and pre-imputation relationship views before any model execution.

\paragraph*{DG2: Enable fair cross-method comparison.}
Method choice should be driven by evidence rather than familiarity. To support this, all models are run under consistent masks and feature handling, and results are summarized with comparable metrics (MAE, RMSE, \(\Delta\)RMSE, runtime) for the current target and masking setting.

\paragraph*{DG3: Support visual plausibility checks.}
Aggregate error alone can hide local artifacts. Linked scatterplots and histograms with stable axis scales let analysts inspect whether imputations preserve relationships, distort tails, or introduce implausible clusters when switching algorithms.

\paragraph*{DG4: Preserve provenance and spatial interpretability.}
For geographically structured data, users must inspect who contributed to an imputed value. The map and donor inspection views therefore expose donor counties and context so analysts can assess neighborhood structure and borrowing plausibility.

Together, these goals frame imputation as an analytic reasoning problem rather than only a predictive task. Users iterate between checking model scores and visually validating whether local behavior matches domain expectations. The interface is therefore designed to keep diagnostics and evidence co-located, reducing context switching and mental overhead across separate tools.

\section{Methodology and System Design}
\paragraph*{Architecture and workflow.}
\tool uses a Python/FastAPI backend and React/TypeScript/D3.js frontend. The backend manages sessions, profiles missingness, runs imputers, and caches results by session/method/target/mask; the frontend links configuration, diagnostics, distribution views, and cross-method comparison panels for rapid iterative analysis (Figure~\ref{fig:workflow_short}, A--G).

\paragraph*{Imputation models.}
We include linear regression, non-spatial kNN, Random Forest, XGBoost, MICE (IterativeImputer with ridge chains), and \gknn. For \gknn, we compute socioeconomic and geographic distance matrices and optimize the blending weight \(\alpha\) and neighborhood size \(k\), using training (non-heldout) data only for each evaluation run.
\begin{equation}
  D_{\text{blend}}=\alpha D_{\text{socio}}+(1-\alpha)D_{\text{geo}}.
\end{equation}

Masked counties are imputed using inverse-distance-weighted averages over the top-\(k\) nearest donors under \(D_{\text{blend}}\), and donor sets are retained for provenance inspection.

\paragraph*{Evaluation.}
For each selected target, we apply a 20\% random holdout mask on observed values and run each algorithm under identical conditions. All model selection and tuning (including gKNN’s $\alpha$ and k) uses training (non-heldout) data only. We report MAE, RMSE, \(\Delta\)RMSE, and runtime in a unified comparison table. In addition to random holdout, we use suppression-like masking (target \(< \tau\), \(\tau = 13\)) and intermediate masking (target within a near-threshold range) to probe more structured forms of missingness. We also compute paired Wilcoxon signed-rank tests~\cite{wilcoxon1945individual} and perform a copula-based uncertainty analysis for \gknn offline~\cite{hollenbach2021multiple, joe2014dependence, nelsen2006introduction}.

\paragraph*{System design rationale.}
The dashboard is organized around iterative analyst behavior rather than a single linear pipeline: users repeatedly switch targets, rerun methods, and revisit comparisons as hypotheses evolve. Caching and synchronized state were therefore prioritized to keep this loop interactive. The design intentionally pairs numeric summaries with linked views so that method rankings can be interpreted with visual evidence, rather than in isolation.

\paragraph*{Backend imputation framework.}
The backend supports uploading, profiling, imputation, evaluation, and export, and caches runs keyed by dataset/target/algorithm/mask to enable responsive cross-method switching with synchronized views and fixed-axis domains.

\subsection{Design process and requirements}
The system design was shaped through iterative feedback from data-analysis users in an academic public-health context. Early sessions revealed fragmented workflows in which missingness diagnosis, model execution, and plausibility checking were handled in separate tools. This motivated a unified interface where diagnosis, benchmarking, and provenance inspection are tightly connected.
Three requirements drove implementation priorities: (R1) algorithmic versatility across spatial and non-spatial imputers; (R2) consistent quantitative evaluation under shared masking settings; and (R3) interpretable outputs that expose donor-level context and cross-method differences. These requirements map directly to the dashboard organization and to backend decisions such as per-configuration caching and standardized metric reporting.



\begin{figure*}[t]
  \centering
  \includegraphics[width=0.9\linewidth]{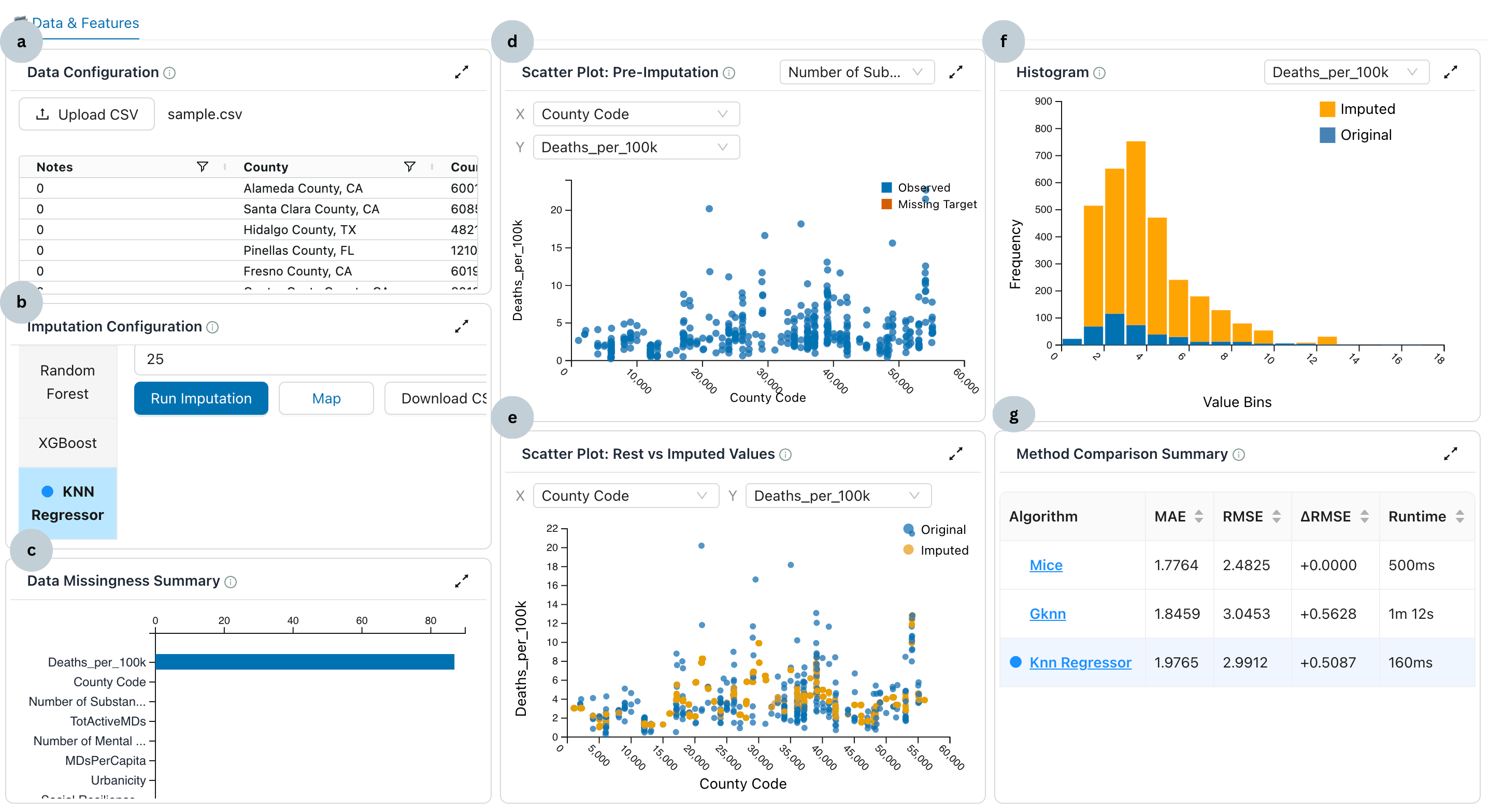}
  \caption{\tool dashboard with coordinated panels (a--g) for data/imputation configuration, missingness diagnosis, pre/post-imputation diagnostics, and cross-method evaluation via the \emph{Method Comparison Summary}.}
  \label{fig:dashboard_short}
\end{figure*}

\section{Frontend Visual Analytics}
The frontend exposes seven coordinated panels (Figure~\ref{fig:dashboard_short}) that align with the end-to-end analyst workflow. Users begin with data and imputation configuration, then inspect missingness and pre-imputation relationships to hypothesize mechanisms (e.g., MCAR/MAR/MNAR). After model execution, linked scatterplots and histograms compare observed versus imputed values under fixed axes, reducing cognitive load when switching methods. The method-comparison panel summarizes MAE, RMSE, \(\Delta\)RMSE, and runtime for all executed methods under the same masking setting.

For spatial datasets, a geomap view supports donor-level provenance. Selecting an imputed county highlights donor counties and exposes their contributions, enabling analysts to assess whether borrowing patterns are geographically and socioeconomically plausible.

\paragraph*{Panel interactions.}
The Data Configuration and Imputation Configuration panels define the active target and method context used throughout the interface. Missingness summaries and pre-imputation views help analysts choose suitable targets and predictors before launching runs. Post-imputation panels then provide complementary checks: scatterplots reveal relationship preservation, histograms reveal marginal shifts, and the comparison table provides rankable quantitative evidence for method choice.
\paragraph*{Cross-view coordination.}
All panels share a synchronized state. Choosing a method in the comparison table immediately updates post-imputation plots and map overlays. This linked behavior supports rapid what-if analysis and reduces context switching between separate tools. In practice, analysts use this coordination to move from high-level ranking to county-level provenance checks without reconfiguration overhead.

\section{Implementation Details}
\paragraph*{Session management and profiling.}
Each upload creates a persistent \texttt{session id} that stores the merged analysis dataframe, feature encoders, and cached algorithm outputs. During ingestion, the backend profiles column roles, null rates, and binary missingness structure to precompute summary views used by the frontend. Co-missingness statistics and missingness--covariate relationships are cached to keep exploratory interactions responsive.

\paragraph*{Robustness.}
The system applies lightweight validation and defensive fallbacks, so exploratory sessions do not fail on imperfect data. String placeholders are normalized to NaN, unmatched geographies are flagged rather than crashing map rendering, and matrix-conditioning issues in chained imputers trigger regularized fallbacks. For \gknn, donor search falls back to broader neighborhoods when strict neighborhoods are unavailable.

\paragraph*{Linked interaction model.}
The interface maintains one active algorithm state shared by table, histogram, scatter, and map views. When analysts switch methods, cached outputs are reused, and axis domains stay stable for direct visual comparison. This design emphasizes comparative reasoning over single-model inspection.

\begin{figure}[t]
  \centering
  \includegraphics[width=\linewidth]{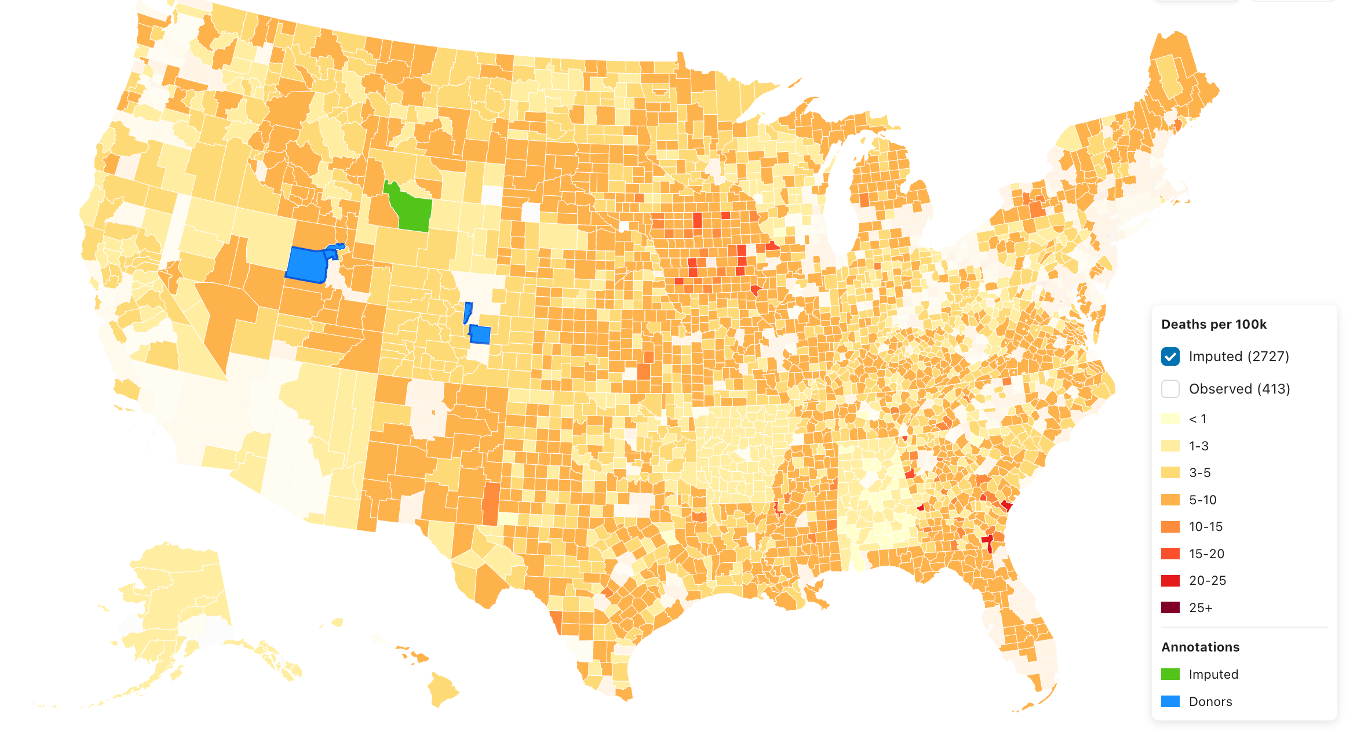}
  \caption{Geomap view for county-level opioid mortality with observed/imputed layer toggles and donor highlighting for selected imputed counties.}
  \label{fig:geomap_short}
\end{figure}

\begin{figure*}[t]
  \centering
  \includegraphics[width=\linewidth]{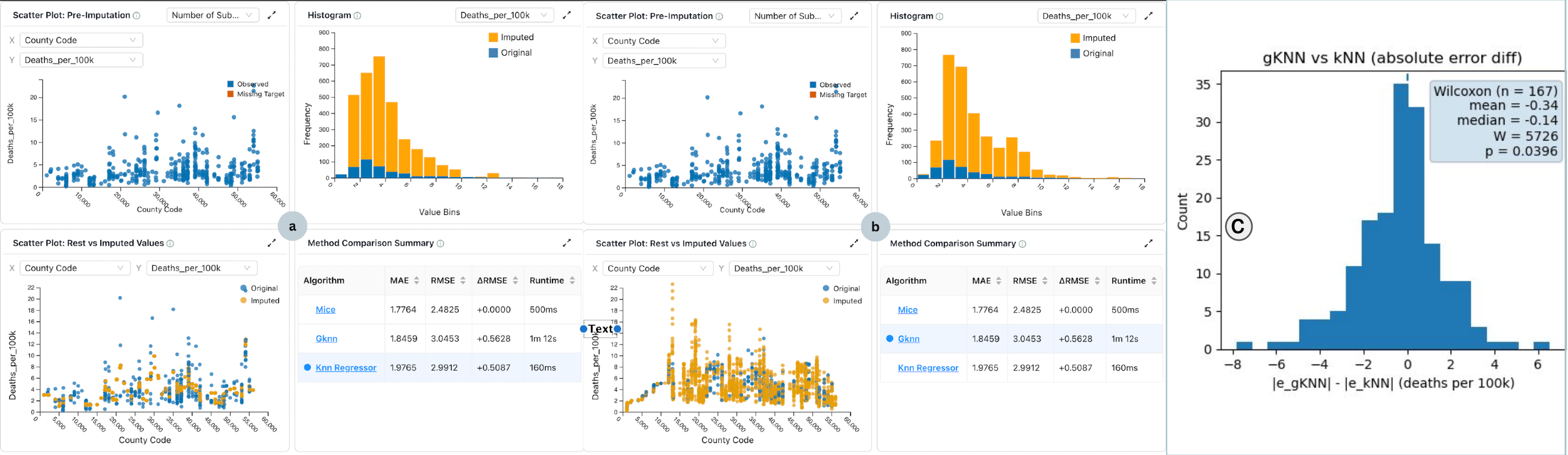}
  \caption{Spatial case study (2010 opioid mortality). Comparison of kNN and \gknn, with distribution of per-county absolute-error differences. Most mass below zero indicates lower error for \gknn.}
  \label{fig:knn_gknn_short}
\end{figure*}

\section{Results}
\begin{table}[t]
  \centering
  \caption{Average MAE (2000--2022) on prescription opioids across suppression-like (Supp.), intermediate (Interm.), random holdout, and pooled Overall settings.}
  \label{tab:metrics}
  \begin{tabular}{@{}lcccc@{}}
    \toprule
    Method & Supp. & Interm. & Random & Overall \\
    \midrule
    Linear Regression & 4.07 & 2.91 & 2.88 & 3.28 \\
    kNN    & 3.86 & 3.18 & 2.81 & 3.28 \\
    Random Forest     & 4.18 & 2.92 & 2.79 & 3.30 \\
    MICE              & 4.05 & 3.06 & 2.88 & 3.33 \\
    XGBoost           & 4.21 & 3.19 & 3.00 & 3.46 \\
    \textbf{\gknn}    & \textbf{3.47} & \textbf{2.76} & \textbf{2.66} & \textbf{2.96} \\
    \bottomrule
  \end{tabular}
\end{table}

Across 23 yearly opioid panels, \gknn achieves the lowest average MAE in all three masking settings (Supp., Interm., Random) (Table~\ref{tab:metrics}), with strongest gains in suppression-like settings where geographic context is most informative. This indicates that integrating spatial proximity with socioeconomic similarity improves robustness when missingness is structurally tied to low-count regions.

We further illustrate method behavior with one spatial case study (opioid mortality) and one non-geographic case (telecom churn), highlighting how \tool supports model comparison and provenance-aware inspection.



\subsection{Spatial opioid mortality: kNN vs gKNN}
In the 2010 opioid case study, paired Wilcoxon tests on per-county absolute errors suggest lower error for \gknn than kNN (\(n=167\), mean difference \(-0.34\), \(p\approx0.0396\)); Figure~\ref{fig:knn_gknn_short} shows the distribution of per-county error differences. Under the same 20\% random holdout mask, kNN often smooths estimates toward broad socioeconomic averages, whereas \gknn better preserves localized variation by blending socioeconomic and geographic neighborhoods. In practice, counties with similar attributes but different regional context are less likely to be treated as equivalent donors in \gknn.

Beyond aggregate error, the difference is operational: analysts can click imputed counties, inspect donor sets, and assess whether borrowing paths are geographically and socioeconomically plausible. For example, we observed cases where kNN achieved competitive error but relied on long-range donors; the donor map made this mismatch visible and motivated preferring gKNN’s more localized donor set. This illustrates why ranking alone is insufficient: provenance provides visual evidence for plausibility in suppression-heavy regions where observed values are sparse.




\subsection{Telecom churn: MICE vs Random Forest}
The telecom churn case provides a complementary non-geographic setting where spatial inductive bias is unnecessary. Using the same 20\% random holdout mask on the continuous \texttt{Offer} target, Random Forest outperforms MICE, and paired Wilcoxon signed-rank tests show a strong separation in per-row absolute errors (\(n=43\), mean difference \(21.9\), \(p\approx2.0\times10^{-9}\)).

This case reinforces \tool's method-agnostic intent: rather than assuming a single best imputer, analysts can identify when flexible non-linear tabular models outperform chained-equation baselines and validate the choice through the same linked diagnostics used in the spatial case.


\section{Evaluation}
\paragraph*{Imputation accuracy}
Across yearly opioid panels and masking settings, \gknn provides the strongest overall error profile in our benchmarked set (Table~\ref{tab:metrics}), with the largest gains in suppression-like conditions where missingness and geography are tightly coupled. This supports the claim that modeling spatial context improves robustness when missingness is structurally tied to place.

Results across datasets also reinforce that no single imputer is universally best: in the non-geographic telecom setting, tree-based models outperform both \gknn and MICE. Accordingly, \tool emphasizes comparison under matched evaluation settings rather than fixed defaults; the method-comparison view combines ranking signals (\(\Delta\)RMSE, MAE/RMSE, runtime) with linked distribution and relationship checks.



\begin{figure}[t]
  \centering
  \includegraphics[width=\linewidth]{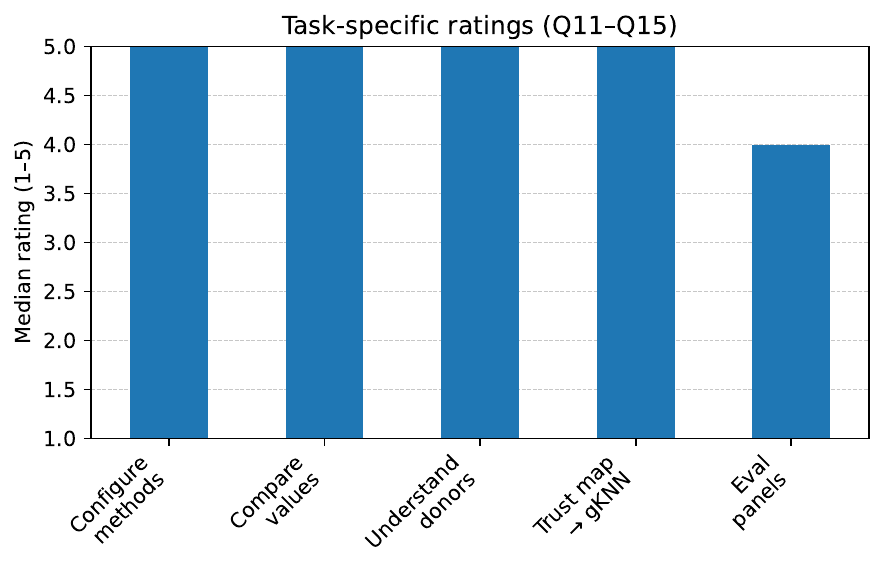}
  \caption{Task-specific usability ratings (Q11--Q15) from the formative study.}
  \label{fig:usability_short}
\end{figure}

\paragraph*{Usability Evaluation}
In a formative usability study (\(N=7\)), the mean SUS score~\cite{brooke1996sus} was 75.4 (SD 15.8), and participants rated method configurability and linked comparisons positively (Q11–Q15 correspond to the workflow tasks in Fig.~\ref{fig:usability_short}). Participants particularly valued fast method switching with immediate visual updates and donor-level provenance in the map workflow; several also reported that stable axes across methods made comparisons easier and reduced re-interpretation effort. Reported pain points centered on runtime latency and presentation polish.




\paragraph*{Uncertainty analysis (summary).}
To complement the holdout error, we use an offline copula-based analysis for \gknn to evaluate interval coverage and width under different conditional sampling assumptions. Coverage remains above 90\% in tested scenarios, while intervals widen under geography-only conditioning, consistent with higher uncertainty when socioeconomic information is reduced. These results motivate future integration of lightweight uncertainty indicators into the dashboard.

\section{Discussion and Conclusion}
\tool combines imputation execution, quantitative benchmarking, and provenance-aware visuals in a single workflow. The results show that spatially informed imputation tends to improve accuracy when geographic autocorrelation is strong, while non-spatial models may dominate in purely tabular settings. This balance is central to the system design: it improves transparency and supports evidence-based model choice rather than algorithm lock-in.

Limitations include the current baseline set, computational cost of richer uncertainty analyses, and scalability constraints at finer geographic resolutions. Future work will extend algorithm coverage, integrate lighter-weight interactive uncertainty cues, and evaluate deployment in operational public-health analysis pipelines.